\documentclass[12pt]{article}
\usepackage{graphicx}
\usepackage{amsmath}
\usepackage{geometry}
\usepackage{hyperref}
\usepackage{booktabs}
\usepackage{float}
\usepackage{subcaption}
\usepackage{caption}
\usepackage{multirow}
\usepackage{longtable}
\usepackage{array}
\usepackage{colortbl}
\usepackage{xcolor}

\usepackage{titling}

\title{Attention Xception UNet (AXUNet): A Novel Combination of CNN and Self-Attention for Brain Tumor Segmentation}

\author{
    Farzan Moodi, MD$^{1,2}$ \\
    Fereshteh Khodadadi Shoushtari, MS$^{1}$ \\
    Gelareh Valizadeh, PhD$^{1}$ \\
    Dornaz Mazinani, BSc$^{1}$ \\
    Hanieh Mobarak Salari, BSc$^{1}$ \\
    Hamidreza Saligheh Rad, PhD$^{1,3}$\thanks{
        Corresponding author. \\
        \textbf{Email:} \texttt{h.salighehrad@qmisg.com}, \texttt{hamid.saligheh@gmail.com} \\
        \textbf{Phone:} +98 (21) 66581505 Ext. 124 \\
        \textbf{Address:} Quantitative Medical Imaging Systems Group, \\
        Research Center for Molecular and Cellular Imaging, \\
        Imam Khomeini Hospital, Keshavarz Boulevard, Tehran, Iran.
    }
}

\date{}  

\pretitle{\begin{center}\LARGE\bfseries}
\posttitle{\vspace{10pt}\end{center}}  % Adds space after title
\preauthor{\begin{center}\large}
\postauthor{\end{center}\vspace{-40pt}}  % Reduce space after authors
\begin{document}

% Title
\maketitle

% Left-aligned affiliations with proper span
\begin{quote}  % Matches abstract width
\small
$^{1}$Quantitative MR Imaging and Spectroscopy Group (QMISG), Tehran University of Medical Sciences, Tehran, Iran.\\
$^{2}$School of Medicine, Iran University of Medical Sciences, Tehran, Iran.\\
$^{3}$Department of Medical Physics and Biomedical Engineering, Tehran University of Medical Sciences, Tehran, Iran.
\end{quote}

% Manually add date only once
\vspace{5pt}  % Reduce space between affiliations and date
\begin{center}
    \small\today
\end{center}
\vspace{5pt}

\begin{abstract}
\noindent Accurate segmentation of glioma brain tumors is crucial for diagnosis and treatment planning. Deep learning techniques offer promising solutions, but optimal model architectures remain under investigation. We used the BraTS 2021 dataset, selecting T1 with contrast enhancement (T1CE), T2, and Fluid-Attenuated Inversion Recovery (FLAIR) sequences for model development. The proposed Attention Xception UNet (AXUNet) architecture integrates an Xception backbone with dot-product self-attention modules, inspired by state-of-the-art (SOTA) large language models such as Google Bard and OpenAI ChatGPT, within a UNet-shaped model. We compared AXUNet with SOTA models. Comparative evaluation on the test set demonstrated improved results over baseline models. Inception-UNet and Xception-UNet achieved mean Dice scores of 90.88 and 93.24, respectively. Attention ResUNet (AResUNet) attained a mean Dice score of 92.80, with the highest score of 84.92 for enhancing tumor (ET) among all models. Attention Gate UNet (AGUNet) yielded a mean Dice score of 90.38. AXUNet outperformed all models with a mean Dice score of 93.73. It demonstrated superior Dice scores across whole tumor (WT) and tumor core (TC) regions, achieving 92.59 for WT, 86.81 for TC, and 84.89 for ET. The integration of the Xception backbone and dot-product self-attention mechanisms in AXUNet showcases enhanced performance in capturing spatial and contextual information. The findings underscore the potential utility of AXUNet in facilitating precise tumor delineation.
\end{abstract}

\bigskip
\section{Introduction}
Gliomas, the most common primary brain tumors, are classified into low-grade (LGG; grades 1–2) and high-grade (HGG; grades 3–4) gliomas \cite{Moodietal2024}. High-grade gliomas carry a poor prognosis, with median survival often under 18 months \cite{Dirvenetal2014}. Various factors influence patient survival, one of which is tumor boundary, as it determines the area for radiation treatment and surgical resection. Therefore, accurate delineation of tumor boundaries is crucial, and this is achieved through a process called tumor segmentation.
Traditionally, tumor segmentation has been performed manually by clinicians. However, the results can vary depending on the clinician's expertise, leading to inconsistent outcomes. Moreover, this manual process can be time-consuming, especially for magnetic resonance imaging (MRI) scans with numerous image slices. Automatic tumor segmentation, on the other hand, can significantly reduce the time required for segmentation. Additionally, by leveraging the knowledge of experts who have delineated tumor regions, automatic segmentation models can transfer their expertise to clinical settings.
Initially, machine learning algorithms comprising supervised or unsupervised algorithms were employed for tumor segmentation. However, these methods had their own challenges. Supervised methods required manual feature extraction, while unsupervised methods, such as clustering or seeding techniques, struggled with determining the initial centroid or seed location, presenting significant hurdles in the segmentation process \cite{Biratuetal2021}.The introduction of the UNet \cite{Ronnebergeretal2015} model revolutionized the field of segmentation, and the Brain Tumor Segmentation Challenge (BraTS) \cite{Menzeetal2015} further drew attention to glioma segmentation.
In this study, building upon previous research, we propose Attention Xception UNet (AXUNet) model that combines the power of UNet architecture with memory-efficient Xception blocks, and an attention module called self-attention that gave rise to the state-of-the-art (SOTA) natural language processing models such as OpenAI chat GPT and Google Bard. Taking inspiration from the multi-attention network (MANet) model \cite{Lietal2022}which has shown successful application in Remote Sensing Images, we endeavored to design, implement, and develop our proposed model. To the best of our knowledge, this is the first utilization of such an algorithm in brain tumor segmentation, which not only marks a novel approach in this field but also yields promising performance.

\section{Related Work}
The UNet model was a pioneering high-performance deep learning-based segmentation model. It featured a distinctive architecture comprising a down sampling encoder and an up sampling decoder, earning its name from the unique U-shaped design \cite{Ronnebergeretal2015}. However, this model lacked some salient characteristics. One key issue was the loss of information through successive layers. The original UNet model could partially solve this by concatenating encoder layers in each depth level with the corresponding decoder layer. However, the problem still continued inside the encoder layers. In other words, information could be lost between the first and last layers of each encoder block. Also, using a deep network with a multitude of layers could lead to extremely low or high gradients that prevent model learning. This downside of UNet led to the introduction of residual blocks in the Resnet model. These blocks are designed to transfer residual input data to the output. The backpropagation mechanism ensures that only necessary data are transformed into the output \cite{Heetal2015}.
Another issue with the UNet model lay in the skip-connection layers. The data from the encoder layers was concatenated with the decoder layers without any filtering or refinement, meaning that both relevant and irrelevant information were transferred, potentially leading to model confusion. While this feature effectively mitigated data loss, it also led to model confusion due to excessive information overload. Eventually, the attention mechanism was introduced \cite{Oktayetal2018}. Attention modules were implemented to make the model concentrate on the most important part of the image which is glioma tumor in our case.
Attention modules can be classified based on four features \cite{Niuetal2021}: 1. The Softness of Attention which is defined as soft (deterministic)/hard (stochastic) or global/local attention. 2. Forms of Input Feature categorized as either item-wise or location-wise, where former comprise of single items like words and latter comprise of inputs that are not discrete like images. 3. Input Representations which can be categorized as distinctive, self (ours), co-attention, hierarchical based on the number of inputs. 4. Output Representations which can
be categorized as single-output, multi-head, and multi-dimensional based on the output of one attention map, multiple attention map for multiple inputs or multiple attention maps for one input, respectively. In the task of segmentation, various attention modules were placed in different parts of UNet. Maji et al. \cite{Majietal2022} proposed an attention gate guided decoder that places an attention gate (AG) between the skip connections and downstream decoder blocks. Jia et al. \cite{Jiaetal2023} took advantage of coordinate attention mechanism where it averages info in three directions i.e. axial, sagittal and coronal using global average pooling in the decoder layer. Zhou et al. \cite{ZhouZhu2023} utilized attention-aware multi-modal fusion module for tumor segmentation. This module consists of multi-sized kernels to capture spatial information and average and global pooling layers to capture channel information. Cao et al. \cite{Caoetal2022} built multiscale contextual attention module combined with residual UNet. The implemented attention module was a Convolutional Block Attention Module (CBAM), designed to extract spatial and channel information through a series of smaller modules primarily utilizing max and average pooling layers. The choice of attention module, based on the specific characteristics provided earlier, can significantly impact the model's performance. We propose AXUNet to address the shortcomings of previous attention-based models by integrating convolutional neural networks (CNNs) with a novel self-attention mechanism.

\section{Materials and Methods}

\subsection{Dataset}
In this study, we utilized BraTS2021 dataset. The BraTS dataset consists of multi-institutional and multi-vendor MRI scans of glioma brain tumors. The dataset includes pre- and post-contrast T1 weighted (T1), T2 weighted (T2), and Fluid-Attenuated Inversion Recovery (FLAIR) sequences. BraTS dataset consists of 1251 LGG and HGG cases. The images were previously co-registered to the same anatomical template and resampled to $1 \mathrm{~mm}^{3}$ voxel resolution. The images had dimensions of $240 \mathrm{X} 240 \mathrm{X} 155$. The corresponding masks were generated through automated segmentation models initially, followed by manual correction by expert radiologists. Tumor regions were segmented into three categories: peritumoral edema (PE), necrotic and non-enhancing tumor (NCR/ECT), and enhancing tumor (ET) \cite{Baidetal2021}.

\subsection{Preprocessing and Data Augmentation}
The image and mask preprocessing involved several steps, as illustrated in Figure 1. These steps aimed to help the model better segment tumor regions.

\subsubsection{Image Preprocessing}
\begin{enumerate}
    \item Sequence Selection and Data Split:
    We tried various combinations of available sequences, and the combination of T1-weighted contrast-enhanced (T1CE), T2, and FLAIR was selected for the final model development. These sequences were then concatenated to create 3-channel image matrices. Regarding the data split, out of the 1251 cases available, we allocated $80 \%$ for training, $10 \%$ for validation, and $10 \%$ for testing.
    \item Tumor ‌Bounding:
    To minimize the influence of irrelevant background data, we restricted the slices to those containing at least $0.7\%$ of tumor tissue.
    \item Cropping:
     To reduce black background with no pixel intensities, the slices were confined to brain boundaries at their highest height and width, resulting in images with dimensions of $128 \times 164$.
    \item Normalization:
    Due to heterogeneous data acquisition parameters across different MRI vendors, the intensity values of MRI images are highly variable. Image normalization adjusts the data to a specific range, ensuring consistent intensity values across all images. For our task, we utilized Min-Max normalization, which scales all images between zero and one. 
    Image normalization was implemented for each channel and image separately, as follows:
    \[
    x_{\text{scaled}} = \frac{x - x_{\min}}{x_{\max} - x_{\min}} \tag{1}
    \]
    where $x$ is a pixel value in the image, and $x_{\min }$ and $x_{\max }$ are the minimum and maximum pixel values in the corresponding channel of the image.
    \item Resizing: In the final step of preprocessing, images were resized to $224 \times 224$.
\end{enumerate}

\begin{figure}[h!]
    \centering
    \includegraphics[width=\linewidth]{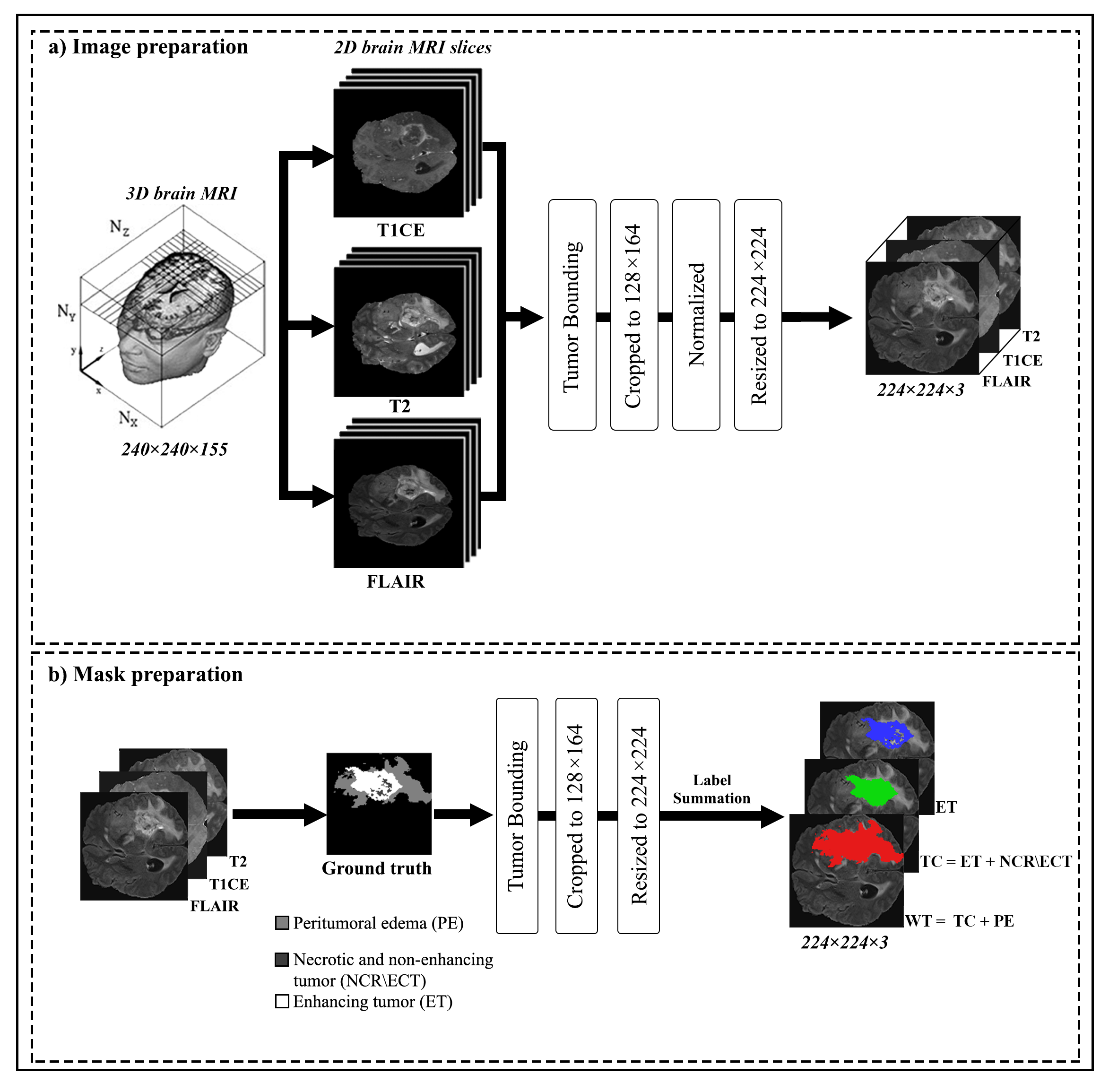}  % Adjust the path and name of the image file as needed
    \captionsetup{justification=raggedright, singlelinecheck=false}  % Left-align the caption
    \caption{Preprocessing workflow: a) Image preparation b) Mask preparation.}  % Add a caption for the image
    \label{fig:image1}  % Optional: Label for referencing the image
\end{figure}

\subsubsection{Mask Preprocessing}
Similar to the preprocessing applied to each image, tumor bounding, cropping, and resizing were first applied to the masks. Then, non-overlapping regions were summed to create a confluent area: NCR/ECT was combined with ET to form the tumor core (TC), TC and PE were summed to form the whole tumor (WT), and ET remained unchanged.

\subsubsection{Data Augmentation}
In order to enhance data variability and mitigate model overfitting, we implemented on-the-fly data augmentation on both images and masks prior to feeding them into the model. To achieve this, we utilized the Albuminations library \cite{Kalinin2018}, enabling us to apply consistent transformations to both images and masks simultaneously. Our augmentation strategy included \textit{randomrotate90}, \textit{horizontal and vertical flips}, and \textit{ShiftScaleRotate} augmentations.

\subsection{Model Architecture}
The overall model architecture is shown in Figure 2(a). The model consists of three main modules, namely the Xception backbone, self-attention module, and decoder blocks (DeBlock).

\subsubsection{Xception Backbone}
Xception, introduced by F. Chollet \cite{Chollet2016}, takes the concept of inception modules and simplifies it by using memory-efficient separable convolution layers. It consists of three sections: the entry flow, middle flow, and exit flow.
The entry flow initiates with two $3 \times 3$ convolution layers followed by ReLU activation. Xception blocks, integral to the entry flow, comprise separable convolution, ReLU, another separable convolution, and max pooling. Additionally, a side branch incorporates a single 1x1 convolution layer, merging its output with that of the max pooling layer (Figure 2(c)). The entry flow concludes with three Xception blocks. The middle flow consists of eight repeated Xception blocks, each containing three consecutive ReLU-activated separable convolution layers. Unlike the entry flow, these blocks omit the 1 x 1 convolution and max pooling layers, instead integrating input with the output of the final separable convolution (Figure 2(d)). The exit flow utilizes the same Xception block as one of the entry flow's Xception blocks, followed by two separable-ReLU layers. We used pretrained weights on the Xception backbone of AXUNet. Utilizing pretrained weights provides a better starting point, resulting in faster convergence and improved outcomes \cite{KhodadadiShoushtari2024, KhodadadiShoushtari2022}.

\begin{figure}[h!]
    \centering
    \includegraphics[width=\linewidth]{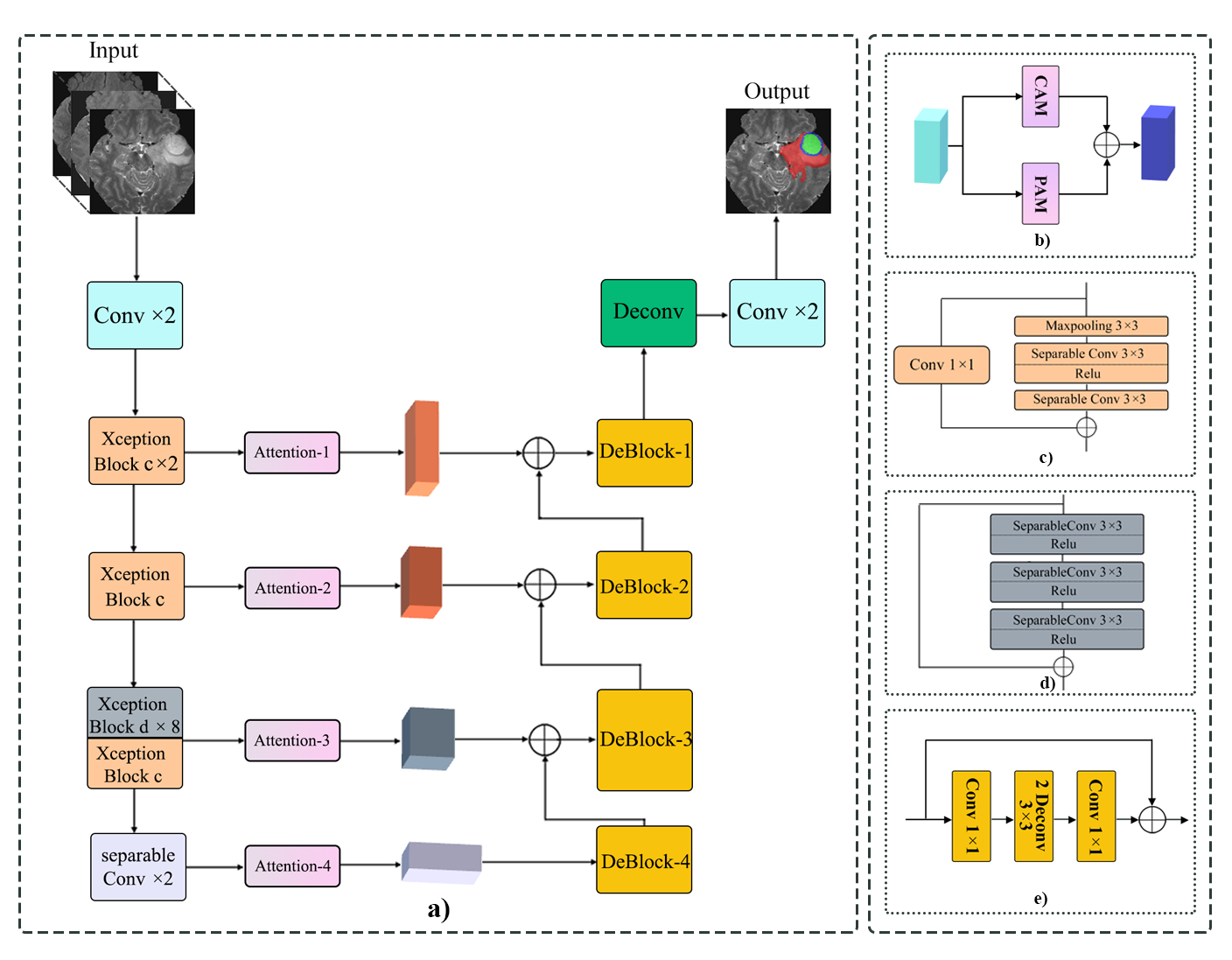}  % Adjust the path and name of the image file as needed
    \captionsetup{justification=raggedright, singlelinecheck=false}  % Left-align the caption
    \caption{Attention Xception UNet (AXUNet): a) Model architecture b) Attention module c) Xception block utilized in the entry and exit flow d) Xception block utilized in the middle flow e) Decoder block. \\
    \vspace{0.5em} % Add space before the note
    Note: The output image was obtained by overlaying the output mask on the input image for better visualization.
    }  % Add a caption for the image
    \label{fig:image2}  % Optional: Label for referencing the image
\end{figure}

\subsubsection{Self-Attention Module}
The self-attention module is a critical component in modern transformer-based architectures \cite{Vaswanietal2017}, pivotal in driving SOTA natural language processing models like OpenAI's ChatGPT and Google's BERT. This module utilizes a scaled dot-product attention mechanism to enable efficient contextual encoding. This module is formed by two key units: the Pixel Attention Module (PAM) and the Channel Attention Module (CAM) (Figure 2(b)).

\subsubsection{Pixel Attention Module (PAM)}
PAM is convolved three times with 1 x 1 conv layer and scaled to produce $1 / 8$ of input features in the output. The three outputs are named query (Q), key (K), and value (V) matrices. Subsequently, Q and K matrices are flattened as follows:
\begin{enumerate}
    \item $\mathrm{Q}(\mathrm{C}, \mathrm{H}, \mathrm{W}) \mathrm{Q}(\mathrm{C}, \mathrm{H} * \mathrm{~W})$
    \item $\mathrm{K}(\mathrm{C}, \mathrm{H}, \mathrm{W}) \mathrm{K}(\mathrm{C}, \mathrm{H} * \mathrm{~W})$
\end{enumerate}
where $\mathrm{C}, \mathrm{H}, \mathrm{W} \in$ Fshow channel, height, and width, respectively. The formula of scaled dot-production attention is as follows:
\[
D(Q, K, V)=\frac{\operatorname{softplast}(Q) \operatorname{softplus}(K)^{T} V}{\operatorname{softplast}(Q) \sum_{j} \operatorname{softplus}(K)_{i, j}^{T} \tag{2}}
\]
This formula calculates the similarity between Q and K using softplast $(Q)$ softplus $(K)^{T}$ equation and divides by similarity of Q with the sum of Ks. This produces a matrix of values with $\mathrm{N} * \mathrm{C}$ dimensions that each row, column value shows the relation of each pair of pixels together. The final result is multiplied to V to apply pixel correlations to the V which is the convolved main image. Finally, the result is summed up with the input image to add the attention map to the image (Figure 3). \\

\begin{figure}[h!]
    \centering
    \includegraphics[width=\linewidth]{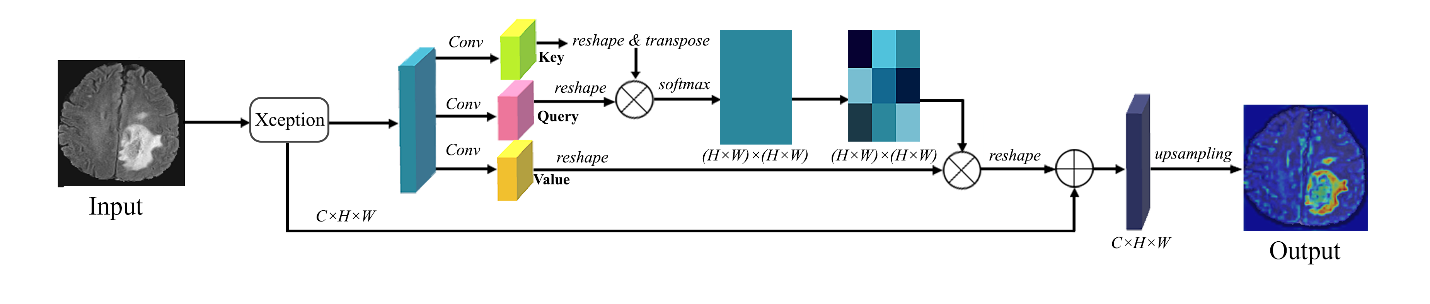}  % Adjust the path and name of the image file as needed
    \captionsetup{justification=raggedright, singlelinecheck=false}  % Left-align the caption
    \caption{Pixel Attention Module (PAM). \\
    \vspace{0.5em} % Add space before the note
    Note: The output image was obtained by generating a Gradient-weighted Class Activation Mapping (Grad-CAM) of the PAM module and overlaying it on the input image for better intuition.
    }  % Add a caption for the image
    \label{fig:image3}  % Corrected label for referencing the image
\end{figure}

\subsubsection{Channel Attention Module (CAM)}
Implementation of CAM module follows the same rules of dot-product attention. The input X (Batch, Channel, Width, Hight) is reshaped three times into $\mathrm{Q}, \mathrm{V}$, and K :
\begin{enumerate}
    \item $\mathrm{Q}, \mathrm{V}=\mathrm{X}$ (Batch, Channel, Width $*$ Hight)
    \item $\mathrm{K}=\mathrm{X}$ (Batch, Width $*$ Hight, Channel)
\end{enumerate}
Using batch matmul, Q and K are multiplied to produce channel attention (batch, channel, channel). Softmax is applied on the attention matrix on the j dimension to weight channels based on their importance. The weighted channel attention is multiplied by V to produce final channel attention map. Finally, channel attention map is added up with the input image (Figure 4). CAM and PAM modules are added together to form self-attention module. In AXUNet, four self-attention modules are placed between the encoder and decoder.

\begin{figure}[h!]
    \centering
    \includegraphics[width=\linewidth]{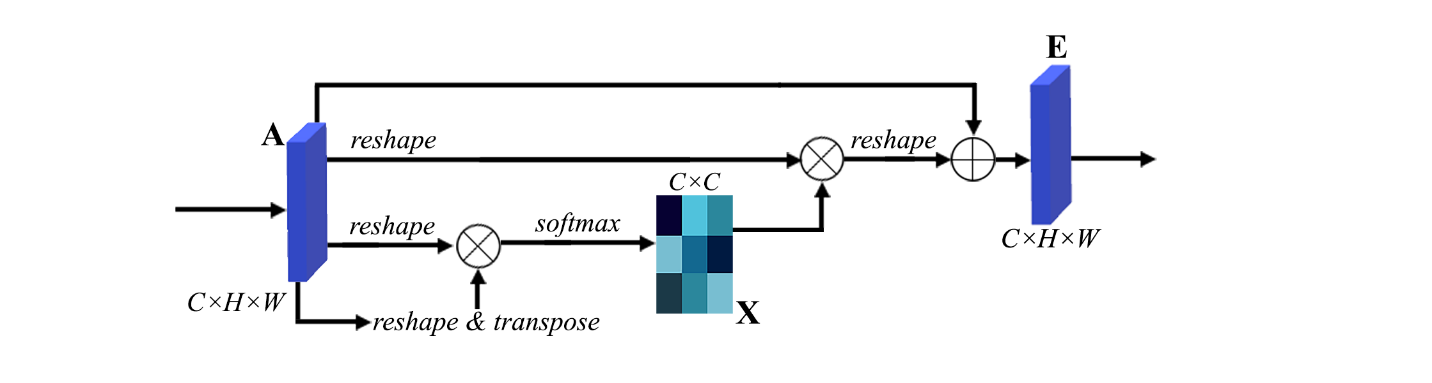}  % Adjust the path and name of the image file as needed
    \captionsetup{justification=raggedright, singlelinecheck=false}  % Left-align the caption
    \caption{Channel Attention Module (CAM).}  % Add a caption for the image
    \label{fig:image4}  % Corrected label for referencing the image
\end{figure}

\subsubsection{DeBlock}
The decoder section consists of four DeBlocks followed by a final deconvolutional layer. Each DeBlock comprises a $1 \times 1$ convolutional layer, two $3 \times 3$ deconvolutional layers, and another $1 \times 1$ convolutional layer, as illustrated in Figure 2(e). The output of each DeBlock is combined with the attention block and passed to the subsequent DeBlock.

\subsection{Loss Function}
The loss function implemented was based on the combination of binary cross entropy (BCE) and Dice loss:
\begin{align}
\operatorname{Loss}_{Dice}(y, \bar{p}) &= 1 - \frac{2 \Sigma(y \bar{p})}{\Sigma y + \Sigma p} \tag{3} \\
\operatorname{Loss}_{BCE}(y, \bar{p}) &= - \Sigma \left[y \log \sigma(x) + (1 - y) \log (1 - \sigma(x))\right] \tag{4} \\
\operatorname{Loss}_{BCE-Dice} &= \operatorname{Loss}_{BCE} + \operatorname{Loss}_{Dice} \tag{5}
\end{align}

where $x$ is the predicted value, $\bar{p}$ is the sigmoid transformed predicted value, and $y$ is the target value. The BCE-Dice loss combines the power of both losses. This loss function was computed independently for each output category, including WT, TC, and ET. Finally, the overall loss was obtained by averaging these individual losses.

\subsection{Model Assessment and Visualization}
To assess to performance of our model, we used Dice score \cite{Dice1945}. Dice score calculates the overlap between the predicted segmentation and ground truth mask using the following formula:
\[
\text { Dice Score }=\frac{2|P \cap G|}{|P|+|G|} \tag{6}
\]
where $P$ and $G$ are predicted and ground truth area, respectively.
For better visualization of intrinsic complexities of the proposed model, we also implemented Gradient-weighted Class Activation Mapping (Grad-CAM) \cite{Selvarajuetal2016}on three layers: the output of last convolution layer of the model, $1^{\text {st }}$ attention module, and DeBlock3's convolution layer.

\subsection{Implementation Details}
We used the PyTorch \cite{Paszkeetal2019} library for model development on a computer system equipped with a 24 GB NVIDIA GTX 3090 GPU and 60 GB of RAM. Through trial and error, we selected the Adam optimizer with an initial learning rate of $10^{-4}$, combined with a cosine annealing scheduler, and a batch size of 64 .
Determining the total number of training epochs involved identifying the epoch where the base model (UNet) achieved its highest Dice score on the validation set. To ensure flexibility, we included a buffer of 10 epochs, resulting in a total of 40 training epochs. Throughout the training process, model weights were saved based on the best performance observed on the validation set. Subsequently, the finalized model underwent evaluation using the unseen test set.

\section{Results}

\subsection{Ablation Study}
We conducted an ablation study to evaluate the contribution of each module in our final model. We first used UNet as the baseline model, achieving a mean Dice score of 93.26, with individual scores of 91.73 for WT, 86.46 for TC, and 84.81 for ET. Replacing the UNet encoder with the Xception backbone slightly reduced the mean Dice score by 0.02. Specifically, the Dice score for TC dropped from 86.46 to 83.74, and for ET from 84.81 to 84.44, while the WT score increased from 91.73 to 92.27. Next, we introduced scaled dot-product attention modules before each skip connection, leading to the final AXUNet model. This approach resulted in a mean Dice score of 93.73. Our proposed AXUNet achieved Dice scores of 92.59 for WT, 86.81 for TC, and 84.89 for ET.

\subsection{Comparison with Base Models}
To thoroughly evaluate our model, we compared it with base models that share similar modular structures and design principles, including Inception-UNet, ResUNet, Attention ResUNet (AResUNet), and Attention Gate UNet (AG-UNet) (Table 1). Inception-UNet achieved a mean Dice score of 90.88, with individual scores of 89.36 for WT, 80.44 for TC, and 82.21 for ET. AResUNet attained the highest mean Dice score of 92.80, with 91.28 for WT, 85.28 for TC, and 84.92 for ET. AG-UNet performed similarly, achieving a mean Dice score of 90.38, with 90.19 for WT, 85.89 for TC, and 83.77 for ET.

\begin{table}[H]
\centering
\captionsetup{width=\linewidth} % Ensures caption width matches table
\caption{Performance evaluation of AXUNet against base models. The highest Dice scores are bolded.}
\begin{tabular}{|l|l|l|l|l|}
\hline
\textbf{Model} & \textbf{Mean Dice Score} & \multicolumn{3}{c|}{\textbf{Regional Dice Score}} \\
\hline
 &  & \textbf{WT} & \textbf{TC} & \textbf{ET} \\
\hline
UNet & 93.26 & 91.73 & 86.46 & 84.81 \\
\hline
Xception-UNet & 93.24 & 92.27 & 83.74 & 84.44 \\
\hline
Inception-UNet & 90.88 & 89.36 & 80.44 & 82.21 \\
\hline
AResUNet & 92.80 & 91.28 & 85.28 & \textbf{84.92} \\
\hline
AG-UNet & 90.38 & 90.19 & 85.89 & 83.77 \\
\hline
AXUNet (proposed) & \textbf{93.73} & \textbf{92.59} & \textbf{86.81} & 84.89 \\
\hline
\end{tabular}
\end{table}

\subsection{Comparative Performance Visualization}
Figure 5 presents the segmentation results of six models alongside the original image and ground truth. WT, ET, and TC regions are depicted in red, blue, and green, respectively. Our proposed AXUNet demonstrates superior performance, particularly in distinguishing closely located points without merging them, as seen in rows 1 and 4 within the ET-TC regions. Conversely, in row 1, AG-UNet abruptly terminates WT segmentation. While all models perform reasonably well in row 2, Inception-UNet creates a bottleneck in the green TC region. In row 3, UNet and Xception-UNet perform well but tend to overestimate the WT region in the posterior edematous area (red). Row 5 highlights AXUNet’s ability to accurately delineate the WT-TC boundary, whereas other models either merge these regions or introduce larger errors. For example, UNet produces a segmentation error in the opposite brain hemisphere, while AG-UNet and Xception-UNet misclassify the area between ET and WT as TC.

\begin{figure}[h!]
    \centering
    \includegraphics[width=\linewidth]{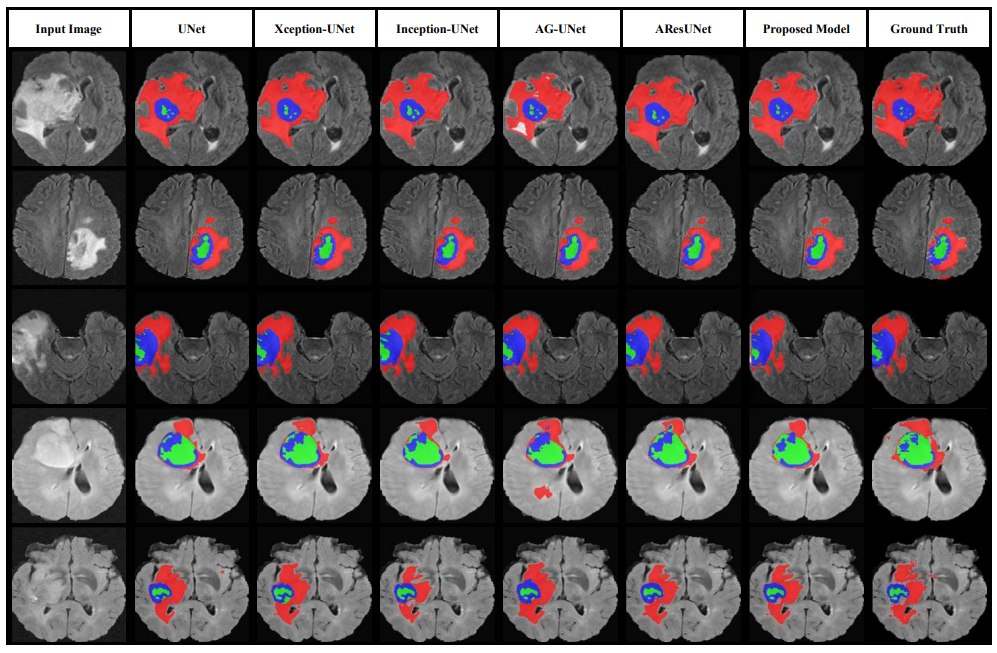}  % Adjust the path and name of the image file as needed
    \captionsetup{justification=raggedright, singlelinecheck=false}  % Left-align the caption
    \caption{Comparative Performance Visualization. Color Legend: WT (Red+ Blue+ Green), ET (Blue), TC (Green+ Blue).}  % Add a caption for the image
    \label{fig:image5}  % Corrected label for referencing the image
\end{figure}

\subsection{Exploring Model Interpretability with Grad-CAM}
Figure 6 presents Grad-CAM visualizations of the last convolution layer, the first attention module, and the convolution layer of DeBlock3 for four different MR images. Areas of higher attention appear in reddish tones, moderate attention in yellow-orange, and the least attention in green. Regions with no attention remain uncolored. Regardless of the segmented region, the model predominantly focuses on the tumor center. In larger regions such as WT, the attention level increases slightly, shifting towards green compared to regions with no attention. An interesting observation arises in Case 4, where deeper layers struggle to capture smaller ET areas, resulting in a clear heatmap. As the analysis moves to shallower layers, the heatmap becomes more pronounced. Additionally, attention is still directed toward brain sulci and cerebrospinal fluid, as indicated by the heatmap of Attention 1. This may introduce attention disturbances due to the high pixel intensities in T2 images.

\begin{figure}[!htbp]  % Allows more flexibility
    \centering
    \includegraphics[width=\linewidth]{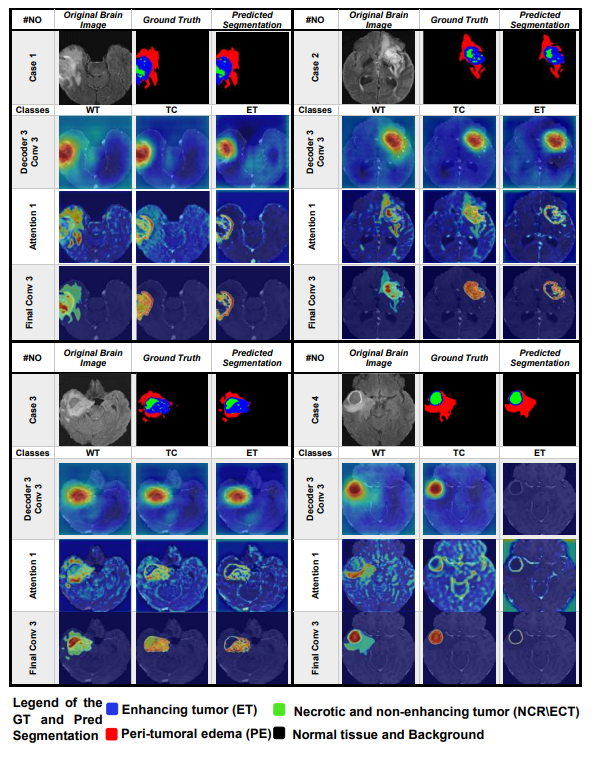}
    \captionsetup{justification=raggedright, singlelinecheck=false}
    \caption{Grad-CAM visualization.}
    \label{fig:image6}
\end{figure}

\section{Discussion}

In this study, we leveraged Xception blocks to mitigate information loss and implemented a scaled dot-product attention mechanism to enhance the model's focus on tumor regions. Overall, our model outperformed conventional segmentation architectures, including UNet, Attention UNet, and Inception UNet, achieving a mean Dice score of 0.937.

The proposed model comprises three main components: a UNet-based architecture, an Xception encoder backbone, and a self-attention module. The UNet structure progresses from shallow to deep layers using max-pooling, capturing image features at varying depths. In the decoder, these features are reconstructed at each corresponding level to ensure accurate image reconstruction. Replacing the UNet encoder with Xception addresses two key limitations of the basic UNet architecture. First, separable convolution layers reduce computational costs compared to conventional convolutions. Second, Xception, originally designed as an extreme version of the Inception model, employs eight blocks of $3 \times 3$ separable convolutions in its middle phase instead of three, enabling more extensive capture of mid-depth features. Additionally, the use of ReLU activation enhances the model’s ability to capture nonlinear features. This improvement was evident in the model’s superior performance compared to InceptionV3. Finally, the self-attention module was integrated to direct the model's focus toward critical regions of the image. While skip connections in the original UNet help counteract information loss during encoding, they do not effectively regulate data transfer, leading to redundant segmentation. By incorporating the self-attention module, our model selectively prioritizes tumor regions, ensuring only relevant image points are transmitted to the decoder.

We compared our model with several state-of-the-art (SOTA) models, considering only studies that used 1,251 cases from the BraTS2021 dataset for a fair comparison. The results are presented in Table 2. Wei et al. \cite{Weietal2022} introduced the High-Resolution Swin Transformer Network (HRSTNet) by integrating transformer blocks from SWINnet into HRNet, achieving Dice scores of 91.90, 87.62, and 82.92 for WT, ET, and TC, respectively. Bukhari et al. \cite{BukhariMohy-ud-Din2021} extended the 3D UNet decoder into three branches, each dedicated to one tumor segment, yielding Dice scores of 92.50 for WT, 89.80 for TC, and 85.60 for ET. Jia et al. \cite{Jiaetal2023} developed a two-branch network with a shared encoder based on 3D UNet. Their decoder incorporated a coordinate attention module and a generative adversarial network for super-resolution image generation, attaining Dice scores of 92.11, 90.09, and 85.13 for WT, TC, and ET, respectively.

Zheng et al. \cite{zheng2023amt} proposed the automated multi-modal Transformer network (AMTNet), incorporating transformer blocks into a UNet structure with parallel encoding via co-learn down-sampling. Their model achieved Dice scores of 92.40, 89.50, and 73.40 for WT, TC, and ET, respectively. Li et al. \cite{li2023transu} introduced TransU${}^{2}$-Net, which integrates transformer blocks within skip connections to capture long-range dependencies, along with a Jump Feature Fusion module in the decoder for high-resolution segmentation, resulting in Dice scores of 92.30, 86.32, and 85.88 for WT, TC, and ET, respectively. Vijay et al. \cite{vijay2023mri} proposed Residual Spatial Pyramid Pooling-powered UNet (SPP-UNet) with attention blocks, where SPP blocks were placed in the skip connections followed by attention modules, achieving Dice scores of 88.70 for WT, 87.90 for TC, and 84.20 for ET.

Ghazouani et al. \cite{ghazouani2024efficient} proposed a brain tumor segmentation model that integrates Swin Transformer and Enhanced Local Self-Attention (ELSA) blocks, utilizing non-overlapping 3D patches processed by a transformer-encoder. A CNN-based decoder with spatial and channel-wise excitation (sSE) refines the extracted features. Their model achieved Dice scores of 91.76 for WT, 88.94 for TC, and 88.95 for ET. Pham et al. \cite{pham2022segtransvae} combined CNNs and transformers within a variational autoencoder (VAE) framework, using CNNs for encoding, transformers in the deepest encoding layer, and a dual-branch decoder with a VAE-inspired regularizer to prevent overfitting. Their model achieved Dice scores of 90.52 for WT, 92.60 for TC, and 85.48 for ET.

As observed in our study and others, transformer-based techniques have improved brain tumor segmentation, enhancing overall performance. While our model achieved the highest Dice score for WT, other models outperformed it in TC and ET segmentation. Comparing different architectures highlights their respective strengths and weaknesses, guiding the development of more refined and effective approaches.

\begin{table}[H]
\centering
\renewcommand{\arraystretch}{1.2} % Adjust row height slightly
\caption{Comparison of the proposed model with several state-of-the-art (SOTA) models on the BraTS 2021 dataset. The highest Dice scores are highlighted in bold.}
\label{tab:comparison}
\begin{tabular}{|p{6cm}|c|c|c|}
\hline
\textbf{Model} & \textbf{WT} & \textbf{TC} & \textbf{ET} \\
\hline
High-Resolution Swin Transformer \cite{Weietal2022} & 91.90 & 82.92 & 87.62 \\
\hline
E1D3 \cite{BukhariMohy-ud-Din2021} & 92.50 & 89.80 & 85.60 \\
\hline
Two-Branch Network with Attention and Super-Resolution Reconstruction \cite{Jiaetal2023} & 92.11 & 90.09 & 85.13 \\
\hline
Automated Multi-Modal Transformer Network (AMTNet) \cite{zheng2023amt} & 92.40 & 89.50 & 73.40 \\
\hline
TransU$^2$-Net \cite{li2023transu}
 & 92.30 & 86.32 & 85.88 \\
\hline
Spatial Pyramid Pooling-Powered 3D UNet \cite{vijay2023mri} & 88.70 & 87.90 & 84.20 \\
\hline
Swin Transformer with Enhanced Local Self-Attention \cite{ghazouani2024efficient} & 91.76 & 88.94 & \textbf{88.95} \\
\hline
SegTransVAE: Hybrid CNN-Transformer \cite{pham2022segtransvae} & 90.52 & \textbf{92.60} & 85.48 \\
\hline
Proposed & \textbf{92.59} & 86.81 & 84.89 \\
\hline
\end{tabular}
\end{table}

\section{Limitations and Future Work}
In our exploration, self-attention module, primarily due to its high computational demands from numerous multiplications. Refining the attention module to reduce these computational burdens is essential. Inspired by separable convolutions, which effectively reduced the computational load of inception modules, similar strategies could be used to streamline the attention mechanism. This approach shows promise in optimizing model efficiency without sacrificing performance. Furthermore, positioning the self-attention module in shallower layers with higher image resolution could improve the model's ability to distinguish regions that are close together. This adjustment would help prevent the generation of confluent areas, leading to more precise segmentation results.

\section{Future Directions and Refinements for BraTS}
The BraTS initiative presents significant potential for advancing surgical practices, yet there remain several areas that could benefit from further refinement in future work. One such area is the delineation of the edema region. As outlined in the BraTS challenge article \cite{li2023brats}, peritumoral edema (PE) consists of both non-enhancing infiltrative tumor and vasogenic edema. In simpler terms, this region includes a mix of tumoral and non-tumoral cells. Refining this segmentation could be achieved by incorporating more advanced MR sequences, such as diffusion- and perfusion-based imaging \cite{scola2023conventional, kazerooni2018characterization}. These enhancements would allow for better differentiation between infiltrative and vasogenic edema, ultimately improving treatment planning by guiding surgical resection and optimizing radiotherapy to minimize damage to healthy tissue.

During our review of the final outputs on the test set, we identified several instances where manual segmentation led to inaccuracies. For example, in the first row of Figure 7, the ground truth image shows two yellow-circled areas within the PE region that extend beyond FLAIR hyperintensities and include brain sulci, which are unrelated to PE. Similarly, in the second row of Figure 7, the manual segmentation encompasses brain sulci and gray matter, leaving a hollow stripe without a clear indication of the actual edema region. While these discrepancies may seem minor, they present two key challenges. First, such inconsistencies could introduce noise during model training. Second, no model would be expected to achieve a perfect Dice score, as it is unlikely to replicate these specific segmentation errors.

Interestingly, our model did not replicate these ground truth inaccuracies and instead produced more precise segmentations. This suggests that the majority of image slices in the dataset were accurately labeled, allowing for effective model training. However, the model's inability to achieve a perfect Dice score on these specific slices highlights the second challenge. These findings may suggest the need for further refinement in the BraTS challenge to enhance segmentation accuracy and overall model performance.

\begin{figure}[H]  % Force the figure to appear exactly here
    \centering
    \includegraphics[width=0.7\linewidth]{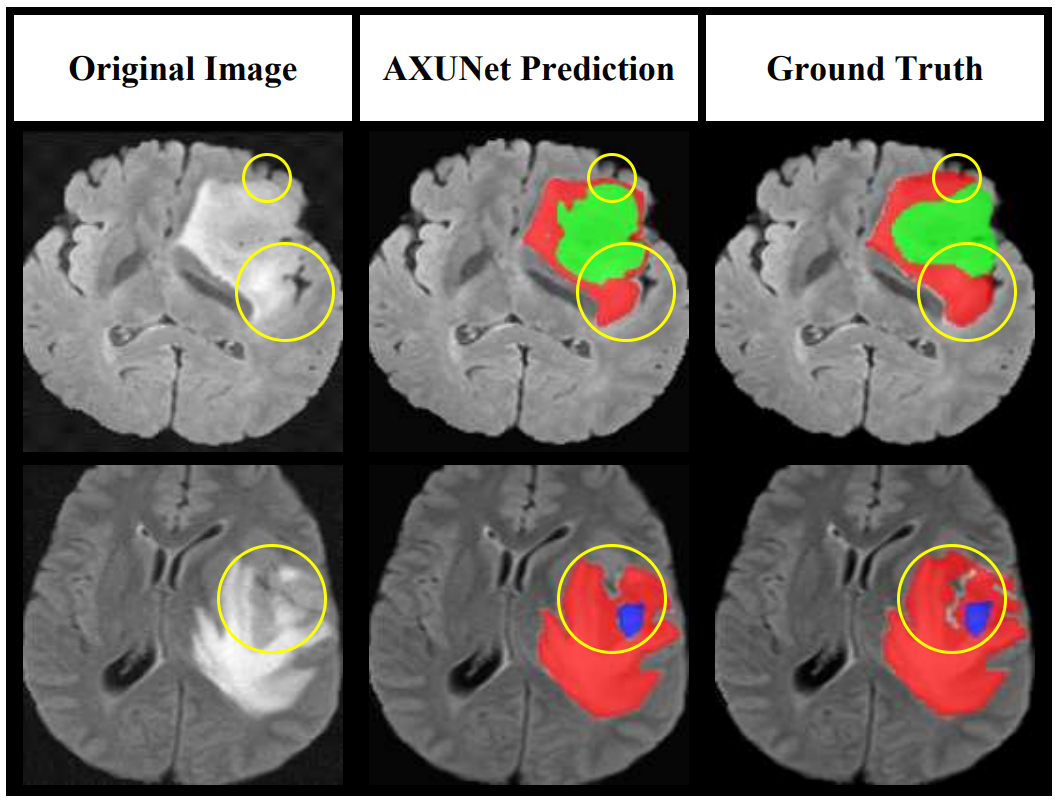}  % Reduce the size by adjusting the width
    \captionsetup{justification=raggedright, singlelinecheck=false}  % Left-align the caption
    \caption{Instances of Areas in Need of Further Refinement. Yellow circles in the ground truth indicate incorrectly segmented areas extending beyond edematous areas and containing brain sulci. Corresponding circles are drawn to represent AXUNet prediction, along with areas in the original image, for comparison.}
    \label{fig:image7}  % Corrected label for referencing the image
\end{figure}

\section{Conclusion}
This study presents AXUNet, a novel framework that combines the robustness of the UNet architecture with the power of convolutional layers, enhanced by Xception blocks for lightweight convolutions through separable convolutions. Furthermore, we improve its performance by incorporating a transformer-based self-attention module, enabling the model to focus attention on tumor regions. While our model shows promising results, further refinement is necessary, including the integration of diverse architectures and rigorous testing in clinical settings to validate its effectiveness in real-world applications.

\section*{Acknowledgements}
This research did not receive any specific grant from funding agencies in the public, commercial, or not-for-profit sectors.

\section*{Code Availability}
Code availability can be provided upon reasonable request. For more information, please contact the corresponding author via email or visit the contact page at \texttt{www.QMISG.com}.

% \section{References}
\bibliographystyle{IEEEtran}  % Choose your preferred style (e.g., plain, ieeetr, unsrt, etc.)

\begin{thebibliography}{10}
\providecommand{\url}[1]{#1}
\csname url@samestyle\endcsname
\providecommand{\newblock}{\relax}
\providecommand{\bibinfo}[2]{#2}
\providecommand{\BIBentrySTDinterwordspacing}{\spaceskip=0pt\relax}
\providecommand{\BIBentryALTinterwordstretchfactor}{4}
\providecommand{\BIBentryALTinterwordspacing}{\spaceskip=\fontdimen2\font plus
\BIBentryALTinterwordstretchfactor\fontdimen3\font minus \fontdimen4\font\relax}
\providecommand{\BIBforeignlanguage}[2]{{%
\expandafter\ifx\csname l@#1\endcsname\relax
\typeout{** WARNING: IEEEtran.bst: No hyphenation pattern has been}%
\typeout{** loaded for the language `#1'. Using the pattern for}%
\typeout{** the default language instead.}%
\else
\language=\csname l@#1\endcsname
\fi
#2}}
\providecommand{\BIBdecl}{\relax}
\BIBdecl

\bibitem{Moodietal2024}
F.~Moodi, F.~Khodadadi~Shoushtari, D.~J. Ghadimi, G.~Valizadeh, E.~Khormali, H.~M. Salari, M.~A.~D. Ohadi, Y.~Nilipour, A.~Jahanbakhshi, and H.~S. Rad, ``Glioma tumor grading using radiomics on conventional mri: A comparative study of who 2021 and who 2016 classification of central nervous tumors,'' \emph{Journal of Magnetic Resonance Imaging: JMRI}, vol.~60, no.~3, pp. 923--938, 2024.

\bibitem{Dirvenetal2014}
L.~Dirven, N.~K. Aaronson, J.~J. Heimans, and M.~J. Taphoorn, ``Health-related quality of life in high-grade glioma patients,'' \emph{Chin J Cancer}, vol.~33, no.~1, pp. 40--45, 2014.

\bibitem{Biratuetal2021}
E.~S. Biratu, F.~Schwenker, Y.~M. Ayano, and T.~G. Debelee, ``A survey of brain tumor segmentation and classification algorithms,'' \emph{Journal of Imaging}, vol.~7, no.~9, p. 179, 2021.

\bibitem{Ronnebergeretal2015}
O.~Ronneberger, P.~Fischer, and T.~Brox, ``U-net: Convolutional networks for biomedical image segmentation,'' \emph{ArXiv}, 2015, abs/1505.04597.

\bibitem{Menzeetal2015}
B.~H. Menze, A.~Jakab, S.~Bauer, and et~al., ``The multimodal brain tumor image segmentation benchmark (brats),'' \emph{IEEE Transactions on Medical Imaging}, vol.~34, no.~10, pp. 1993--2024, 2015.

\bibitem{Lietal2022}
R.~Li, S.~Zheng, C.~Zhang, and et~al., ``Multiattention network for semantic segmentation of fine-resolution remote sensing images,'' \emph{IEEE Transactions on Geoscience and Remote Sensing}, vol.~60, pp. 1--13, 2022.

\bibitem{Heetal2015}
K.~He, X.~Zhang, S.~Ren, and J.~Sun, ``Deep residual learning for image recognition,'' \emph{2016 IEEE Conference on Computer Vision and Pattern Recognition (CVPR)}, pp. 770--778, 2015.

\bibitem{Oktayetal2018}
O.~Oktay, J.~Schlemper, L.~L. Folgoc, and et~al., ``Attention u-net: Learning where to look for the pancreas,'' \emph{ArXiv}, 2018, abs/1804.03999.

\bibitem{Niuetal2021}
Z.~Niu, G.~Zhong, and H.~Yu, ``A review on the attention mechanism of deep learning,'' \emph{Neurocomputing}, vol. 452, pp. 48--62, 2021.

\bibitem{Majietal2022}
D.~Maji, P.~Sigedar, and M.~Singh, ``Attention res-unet with guided decoder for semantic segmentation of brain tumors,'' \emph{Biomedical Signal Processing and Control}, vol.~71, p. 103077, 2022.

\bibitem{Jiaetal2023}
Z.~Jia, H.~Zhu, J.~Zhu, and P.~Ma, ``Two-branch network for brain tumor segmentation using attention mechanism and super-resolution reconstruction,'' \emph{Computers in Biology and Medicine}, vol. 157, p. 106751, 2023.

\bibitem{ZhouZhu2023}
T.~Zhou and S.~Zhu, ``Uncertainty quantification and attention-aware fusion guided multi-modal mr brain tumor segmentation,'' \emph{Comput Biol Med}, vol. 163, p. 107142, 2023.

\bibitem{Caoetal2022}
T.~Cao, G.~Wang, L.~Ren, Y.~Li, and H.~Wang, ``Brain tumor magnetic resonance image segmentation by a multiscale contextual attention module combined with a deep residual unet (mca-resunet),'' \emph{Physics in Medicine \& Biology}, vol.~67, no.~9, p. 095007, 2022.

\bibitem{Baidetal2021}
U.~Baid, S.~Ghodasara, M.~Bilello, and et~al., ``The rsna-asnr-miccai brats 2021 benchmark on brain tumor segmentation and radiogenomic classification,'' \emph{ArXiv}, 2021, abs/2107.02314.

\bibitem{Kalinin2018}
A.~A. Kalinin, ``Albumentations: fast and flexible image augmentations,'' 2018, arXiv e-prints.

\bibitem{Chollet2016}
F.~Chollet, ``Xception: Deep learning with depthwise separable convolutions,'' \emph{2017 IEEE Conference on Computer Vision and Pattern Recognition (CVPR)}, pp. 1800--1807, 2016.

\bibitem{KhodadadiShoushtari2024}
F.~Khodadadi~Shoushtari, A.~N.~V. Dehkordi, and S.~Sina, ``Quantitative and visual analysis of data augmentation and hyperparameter optimization in deep learning-based segmentation of low-grade glioma tumors using grad-cam,'' \emph{Annals of Biomedical Engineering}, vol.~52, no.~5, pp. 1359--1377, 2024.

\bibitem{KhodadadiShoushtari2022}
F.~Khodadadi~Shoushtari, S.~Sina, and A.~N.~V. Dehkordi, ``Automatic segmentation of glioblastoma multiform brain tumor in mri images: Using deeplabv3+ with pre-trained resnet18 weights,'' \emph{Phys Med}, vol. 100, pp. 51--63, 2022.

\bibitem{Vaswanietal2017}
A.~Vaswani, N.~M. Shazeer, N.~Parmar, and et~al., ``Attention is all you need,'' 2017.

\bibitem{Dice1945}
L.~R. Dice, ``Measures of the amount of ecologic association between species,'' \emph{Ecology}, vol.~26, no.~3, pp. 297--302, 1945.

\bibitem{Selvarajuetal2016}
R.~R. Selvaraju, A.~Das, R.~Vedantam, and et~al., ``Grad-cam: Visual explanations from deep networks via gradient-based localization,'' \emph{International Journal of Computer Vision}, vol. 128, no.~2, pp. 336--359, 2016.

\bibitem{Paszkeetal2019}
A.~Paszke, S.~Gross, F.~Massa, and et~al., ``Pytorch: An imperative style, high-performance deep learning library,'' 2019, abs/1912.01703.

\bibitem{Weietal2022}
C.~Wei, S.~Ren, K.~Guo, H.~Hu, and J.~Liang, ``High-resolution swin transformer for automatic medical image segmentation,'' \emph{Sensors (Basel, Switzerland)}, vol.~23, 2022.

\bibitem{BukhariMohy-ud-Din2021}
S.~T. Bukhari and H.~Mohy-ud Din, ``E1d3 u-net for brain tumor segmentation: Submission to the rsna-asnr-miccai brats 2021 challenge,'' 2021.

\bibitem{zheng2023amt}
S.~Zheng, J.~Tan, C.~Jiang, and L.~Li, ``Automated multi-modal transformer network (amtnet) for 3d medical images segmentation,'' \emph{Phys. Med. Biol.}, vol.~68, no.~2, p. 025014, 2023, published online 2023/01/09.

\bibitem{li2023transu}
X.~Li, X.~Fang, G.~Yang, S.~Su, L.~Zhu, and Z.~Yu, ``Transu²-net: An effective medical image segmentation framework based on transformer and u²-net,'' \emph{IEEE J. Transl. Eng. Health Med.}, vol.~11, pp. 441--450, 2023.

\bibitem{vijay2023mri}
S.~Vijay, T.~Guhan, K.~Srinivasan, P.~Vincent, and C.~Y. Chang, ``Mri brain tumor segmentation using residual spatial pyramid pooling-powered 3d u-net,'' \emph{Front. Public Health}, vol.~11, p. 1091850, 2023.

\bibitem{ghazouani2024efficient}
F.~Ghazouani, P.~Vera, and S.~Ruan, ``Efficient brain tumor segmentation using swin transformer and enhanced local self-attention,'' \emph{International Journal of Computer Assisted Radiology and Surgery (Int J CARS)}, vol.~19, pp. 273--281, 2024.

\bibitem{pham2022segtransvae}
Q.~Pham, H.~Nguyen-Truong, N.~Phuong, and et~al., ``Segtransvae: Hybrid cnn - transformer with regularization for medical image segmentation,'' \emph{Preprint}, pp. 1--5, 2022.

\bibitem{li2023brats}
H.~Li, G.~Conte, S.~Anwar, and et~al., ``The brain tumor segmentation (brats) challenge 2023: Brain mr image synthesis for tumor segmentation (brasyn),'' \emph{ArXiv}, Jun 2023.

\bibitem{scola2023conventional}
E.~Scola, G.~Del~Vecchio, G.~Busto, and et~al., ``Conventional and advanced magnetic resonance imaging assessment of non-enhancing peritumoral area in brain tumor,'' \emph{Cancers}, vol.~15, no.~11, p. 2992, 2023.

\bibitem{kazerooni2018characterization}
A.~Fathi~Kazerooni, M.~Nabil, M.~Zeinali~Zadeh, and et~al., ``Characterization of active and infiltrative tumorous subregions from normal tissue in brain gliomas using multiparametric mri,'' \emph{Journal of Magnetic Resonance Imaging}, vol.~48, no.~4, pp. 938--950, 2018.

\end{thebibliography}

  % This should match the name of your .bib file (no need to add .bib extension)

\end{document}